\documentstyle[epsfig,12pt]{test}

\newcommand{\klgg}    {\mbox{$K^\circ_L \! \rightarrow \!  \gamma\gamma$ }}
\newcommand{\klpp}    {\mbox{$K^\circ_L \! \rightarrow \! \pi^+ \pi^-$ }}
\newcommand{\klmm}    {\mbox{$K^\circ_L \! \rightarrow \! \mu^+ \mu^-$ }}
\newcommand{\klee}    {\mbox{$K^\circ_L \! \rightarrow \! e^+ e^-$ }}
\newcommand{\klme}    {\mbox{$K^\circ_L \! \rightarrow \! \mu e$ }}
\newcommand{\kpme}    {\mbox{$K^+ \! \rightarrow \! \pi^+ \mu^+ e^-$ }}
\newcommand{\klpme}   {\mbox{$K^\circ_L \! \rightarrow \! \pi^\circ \mu e$ }}
\newcommand{\klpnn}   {\mbox{$K^\circ_L \! \rightarrow \! \pi^\circ \nu \overline{\nu}$ }}
\newcommand{\kzpnn}    {\mbox{$K \! \rightarrow \! \pi \nu \overline{\nu}$ }}
\newcommand{\klpll}   {\mbox{$K^\circ_L \! \rightarrow \! \pi^\circ \ell^+ \ell^-$ }}
\newcommand{\kspee}   {\mbox{$K^\circ_S \! \rightarrow \! \pi^\circ e^+ e^-$ }}
\newcommand{\klpee}   {\mbox{$K^\circ_L \! \rightarrow \! \pi^\circ e^+ e^-$ }}
\newcommand{\klpmm}   {\mbox{$K^\circ_L \! \rightarrow \! \pi^\circ \mu^+ \mu^-$ }}
\newcommand{\kpnn}    {\mbox{$K^+ \! \rightarrow \! \pi^+ \nu \overline{\nu}$ }}

\newcommand{\klllgg}  {\mbox{$K^\circ_L \! \rightarrow \! \ell^+ \ell^- \gamma\gamma$ }}
\newcommand{\klpgg}   {\mbox{$K^\circ_L \! \rightarrow \! \pi^\circ \gamma\gamma$ }}

\newcommand{\bpsiks}  {\mbox{$B^\circ_d \! \rightarrow \! \psi K^\circ_S$ }}
\newcommand{\bsbd}    {\mbox{$B_s - \overline{B}_s$ }}
\newcommand{\kpen}    {\mbox{$K^+\!\rightarrow\!\pi^\circ e^+\nu_e$ }}
\newcommand{\kpp}     {\mbox{$K^+ \! \rightarrow \! \pi^+ \pi^\circ$ }}

\newcommand{\vtd}     {\mbox{$V_{td}$}}

\begin{document}
\title{ 
\vspace{-1.5cm}
\rightline{\small\rm BNL--67589}
\vspace{-0.3cm}
\rightline{\small\rm July 1, 2000}
\vspace{1cm}
Rare Kaon Decay Experiments
\footnote{ To be published in the {\it Proceedings of the Workshop on Strange 
Quarks in Hadrons, Nuclei, and Nuclear Matter; Athens, Ohio, May 12-13, 2000};
Ed. K.~Hicks} }
\author{ Steve Kettell \\
\small\em Brookhaven National Laboratory\\ 
\small\em Upton, NY 11973}
\date{}

\maketitle

\abstract{ The current status of rare kaon decay experiments is
reviewed.  A large number of new results are available from
several very sensitive experiments at BNL, FNAL and CERN.
New limits in the search for Lepton Flavor Violation are
discussed, as are new measurements of the CKM matrix.}

\section{Introduction}

The study of rare kaon decays has played a key role in the development
of the standard model (SM), and the field continues to have
significant impact.  Several recent reviews of the field are or will be soon
available~\cite{review2000}.  The two areas of greatest import are the
determination of fundamental standard-model parameters, such as
CKM~\cite{kobayashi} mixing and {\it CP} violation, and the search for
physics beyond the standard model (BSM) through the search for lepton
flavor violating (LFV) decays.

\section{Lepton Flavor Violating Decays}

There is solid experimental evidence for the exact conservation of an
additive quantum number for each family of charged leptons.  While
there is no SM mechanism for LFV (non-zero m$_{\nu}$ induces LFV in
the charged lepton sector at a level that is too small to observe),
there is no underlying gauge symmetry preserving lepton flavor; and
many extensions to the SM predict LFV.  Observation of LFV would be
unambiguous evidence for physics beyond the SM.

Due to the relatively long kaon lifetime, copious production at fixed
target proton accelerators, and very sophisticated experimental
techniques, the mass scale probed by rare kaon decay experiments is
quite high.  This can be seen by comparing the \klme decay, through a
hypothetical LFV vector boson with coupling $g_X$ and mass $M_X$ to
the conventional $K_{\mu2}$ decay ($g$ and $M_W$): $M_X > 200 \;{\rm
TeV/c^2} \times g_X/g \times \left[ 10^{-12}/ \rm
B(\klme)\right]^{1/4}.$ For current experimental sensitivities at the
level of $10^{-12}$, mass scales in excess of 100 TeV are explored
(for the usual electroweak coupling).

\begin{figure}[ht]
\epsfig{file=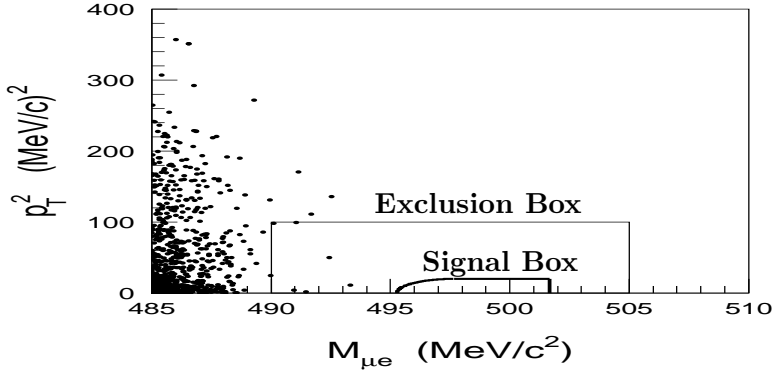,height=1.75in,width=4.25in,angle=0} 

\vspace{-1.85cm}
\hbox{\bf \hspace{2.in}  Exclusion Box}
\vspace{0.3cm}
\hbox{\bf \hspace{2.25in}  Signal Box}
\vspace{1.05cm}

\caption{Final E871 data sample after all cuts, with no events in the
signal region. The exclusion box was used to set cuts in an unbiased
way on data far from the signal region. The shape of the signal box
was optimized to maximize signal/background.
\label{fig:e871}}
\end{figure}

The E871 experiment at BNL, a search for \klme, has been completed,
with two long runs during 1995--96. The E871 analysis of the data set
used a `blind analysis' technique, in which selection criteria were
devised and backgrounds measured on the data outside of an exclusion
region. These selection criteria were then applied to the remaining
data and background measurements were compared to the actual number of
events.  No events were seen in the signal region (see
Fig.~\ref{fig:e871}), with an expected background of 0.1 events
and the 90\% CL limit on this decay is B(\klme) $<
4.7\times10^{-12}$~\cite{e871_me}.  There are no plans to
pursue this decay further.

The E865 experiment at BNL, designed to search for the decay \kpme, an
analog to \klme which is sensitive to LFV interactions with different
quantum numbers, collected data during 1995, 1996 and 1998. The 90\%
CL limits on this mode from the 1995 and 1996 runs are B(\kpme)
$< 2.1\times10^{-10}$ and $< 3.9\times10^{-11}$ respectively.
Combining these results with that of the predecessor experiment E777,
the 90\% CL limit is B(\kpme) $< 2.8\times10^{-11}$~\cite{e865_pme}.  The final
sensitivity, including 1998 data, is expected to be $\sim$3 times
better.  There are no plans to continue this search.

The current 90\% CL limit for \klpme is from E799-I~\cite{e799_pme} at
FNAL, with B(\klpme) $< 3.1\times10^{-9}$. This measurement has very
little background, so E799-II (KTeV) will be able to substantially
improve upon this limit, reaching sensitivities close to E865.

\boldmath
\section{CKM Matrix}
\unboldmath

The \kzpnn modes: \kpnn and \klpnn, are the `golden modes' for
determining CKM matrix parameters.  Both of these modes can be very
precisely calculated from fundamental SM parameters.  The \kpnn mode
is sensitive to the magnitude of the poorly known $\lambda_t\equiv
V_{ts}^*V_{td}$ and \klpnn is purely direct-{\it CP}-violating and
sensitive to $Im(\lambda_t)$.

Two other modes for which it may be possible to extract fundamental
CKM parameters are \klmm and \klpll.  However, in both cases, large
long-distance contributions limit the usefulness of these modes.
Additional measurements of
some radiative kaon decays, as well as chiral perturbation theory
(ChPT) work, are needed to extract the short distance
physics~\cite{dambrosio,dumm,valencia,derafael,donoghue}.  In the case of \klpee and
\klpmm, there are also significant backgrounds.

\boldmath
\subsection{\klmm and \klpll}
\unboldmath

The mode \klmm, whose small rate has played such an important role in
the development of the SM (e.g. the GIM mechanism and the prediction
of the charm quark~\cite{glashow}), has now been measured (relative to
\klpp) to the unprecedented precision of 1.5\%, with 6200 events
observed by E871~\cite{e871_mm}.

The decay \klmm is dominated by \klgg with the two real photons
converting to a $\mu^+$$\mu^-$ pair. This contribution is precisely
calculated using QED from a measurement of the \klgg branching
ratio. There is also a long distance dispersive contribution through
off-shell photons, which has been calculated~\cite{dambrosio,dumm}
although there is some dispute as to the reliability of these
calculations~\cite{valencia,derafael}.  Most interesting is the short
distance contribution, through internal quark loops, dominated by the
top quark.  A measurement of this short distance contribution is
sensitive to the real part of the $\lambda_t$ or, equivalently, the
Wolfenstein~\cite{wolfenstein} parameter $\rho$~\cite{buras2}:
\begin{equation}
B_{SD}(\klmm) = 6.0 \times 10^{-3} [Re(\lambda_t)-6.7\times10^{-5}]^2 \sim 9\times10^{-10}
\end{equation}
The current measurement of the branching
ratio B(\klmm) $= (7.18 \pm 0.17) \times 10^{-9}$ by the E871
collaboration~\cite{e871_mm} represents a factor of three improvement.
This value is only slightly above the unitarity bound from the
on-shell two photon contribution
of B$_{abs}$(\klmm) = $(7.07\pm0.18) \times 10^{-9}$
and leaves very little room for a short distance contribution.
Using estimates of the long distance dispersive
contribution~\cite{dambrosio}, a 90\% CL limit of $\rho >-0.33$ 
is obtained~\cite{e871_mm}.

The decay \klee is predominantly through two off-shell
photons, making this decay less interesting for
extracting SM parameters. However, the recent observation of four
events by E871~\cite{e871_ee}, with B(\klee) $= (8.7^{+5.7}_{-4.1})
\times 10^{-12}$ is consistent with 
ChPT predictions~\cite{dumm,valencia} and is the smallest branching
ratio ever measured for any elementary particle decay.

The \klpll modes can proceed via the direct-{\it CP}-violating processes. This
short-distance contribution is given by~\cite{buras2}
\begin{equation}
B_{SD}(\klpee) = 2.5 \times 10^{-4} [Im(\lambda_t)]^2 \sim 5\times 10^{-12}.
\end{equation}
The muon mode, \klpmm, is expected to be five times smaller.
Unfortunately, the decay \klpee can occur in two other ways: an
indirect-{\it CP}-violating contribution (which can be determined from
measurement of \kspee) and a {\it CP}-conserving contribution (which
may be calculated from the \klpgg rate at low invariant $\gamma\gamma$
mass). These contributions may be comparable to or larger than the
direct-{\it CP}-violating contribution.  Even more formidable is the
background from \klllgg, as pointed out by Greenlee~\cite{greenlee}.
The KTeV experiment at FNAL, analyzing data from 1997, has
significantly improved the limit on \klpee.  Two events, consistent
with the expected background of 1.1 events, were found in the signal
region~\cite{e799_pee}, giving a 90\% CL limit of B(\klpee) $ <
5.6\times10^{-10}$. A similar analysis of the related muon mode
resulted in two events, consistent with the expected background of $0.9\pm0.2$
events, in the signal region~\cite{e799_pmm}, and a slightly
smaller upper limit, $B(\klpmm) < 3.4\times 10^{-10} \; \; (90\%\,{\rm
CL}) $.  Improvement in limits on both of these modes will be slow due
to the presence of background.

There is a wealth of other new measurements in the kaon system
reported recently from BNL, FNAL and CERN~\cite{review2000,daphne}.
Many of these are useful for understanding long-distance effects in
\klmm or \klpll and for determining the background to \klpll.  These
measurements are substantially improved over previous values and even
larger improvements will be obtained when the complete data sets are
analyzed.

\subsection{Golden Modes}

The \kpnn and \klpnn decays are the `golden modes' for measuring CKM
parameters; and, along with the other golden mode \bpsiks and perhaps
the ratio of the  mixing frequencies of $B_s$ and $B_d$ mesons, 
provide the best opportunity to over-constrain the
unitary triangle and to search for new physics.  The most powerful
tests of our understanding of CP-violation and quark mixing will come
from comparison of the results from B meson and kaon decays with
little theoretical ambiguity. The two premier tests are expected to
be:
\begin{itemize}
  \item Comparison of the angle $\beta$ from the ratio
B(\klpnn)/B(\kpnn) and the CP asymmetry in the decay
\bpsiks~\cite{sinb,grossman}.  
  \item Comparison of the magnitude $|\vtd|$ from \kpnn and the ratio of 
the mixing frequencies of $B_s$ to $B_d$ mesons~\cite{bb3}.
\end{itemize}

The unitarity of the CKM matrix can be expressed as
\begin{equation}
V^*_{us}V_{ud} + V^*_{cs}V_{cd} + V^*_{ts}V_{td} = 
\lambda_u + \lambda_c + \lambda_t = 0
\end{equation}
with the three vectors $\lambda_i\equiv V^*_{is}V_{id}$ converging to
form a very elongated triangle in the complex plane. The first vector
$\lambda_u=V^*_{us}V_{ud}$ is well determined from the decay \kpen
($K_{e3}$).  The height can be measured by \klpnn and the third side
$\lambda_t=V^*_{ts}V_{td}$ will be measured by \kpnn.  The decay
\klpnn offers the best opportunity for measuring the Jarlskog
invariant $J_{CP}$~\cite{jarlskog}.  The \kzpnn decays are sensitive
to the magnitude and imaginary part of $\lambda_t$.  Measurements of
these two modes, along with $K_{e3}$, will then completely determine
the unitarity triangle.

The theoretical uncertainty in \kpnn ($\sim$7\%) is small and even
smaller in \klpnn ($\sim$2\%); in both cases the hadronic matrix
element is extracted from B($K_{e3}$).  The branching ratios have been
calculated to next-to-leading-log approximation~\cite{bb1}, complete
with isospin violation corrections~\cite{marciano} and
two-loop-electroweak effects~\cite{bb2}. Based on current
understandings of SM parameters, the branching ratios can be expressed
as~\cite{bb3}:
\begin{eqnarray}
B(\kpnn) & = &\frac{\kappa_+ \alpha^2 B(K_{e3})}
        {2 \pi^2 \sin^4 \theta_W |V_{us}|^2} 
      \sum_l  |X_t\lambda_t + X_c^l\lambda_c|^2\\ \nonumber
 & = & 3.6\times10^{-4}([Re(\lambda_t)-1.4\times10^{-4}]^2 + 
[Im(\lambda_t)]^2) \\ \nonumber
         & = & (0.82\pm0.32)\times10^{-10} \\
B(\klpnn) & = & \frac{\tau_{K_L}}{\tau_{K^+}}
\frac{\kappa_L\alpha^2 B(K_{e3})}{2\pi^2sin^4\theta_W |V_{us}|^2}
\sum_{l} |Im(\lambda_t)X_t|^2 \\ \nonumber
 & = & 1.6\times10^{-3}[Im(\lambda_t)]^2 = (3.1\pm1.3)\times10^{-11}, 
\end{eqnarray}
where $\kappa$ are the isospin corrections and the Inami-Lim
functions~\cite{buras}, $X_q$, are functions of $x_q \equiv$
$M_q^2$/$M_W^2$ where $M_q$ is the mass of the quark $q;$ these
contain QCD corrections.  In addition, it is possible to place a
theoretically unambiguous upper limit on \kpnn from the current limit
on \bsbd mixing, B(\kpnn) $< 1.67\times10^{-10}$~\cite{bb3}.
 
\boldmath
\subsubsection{\kpnn}
\unboldmath

\begin{figure}[ht]
\hspace{1.cm}
\epsfig{file=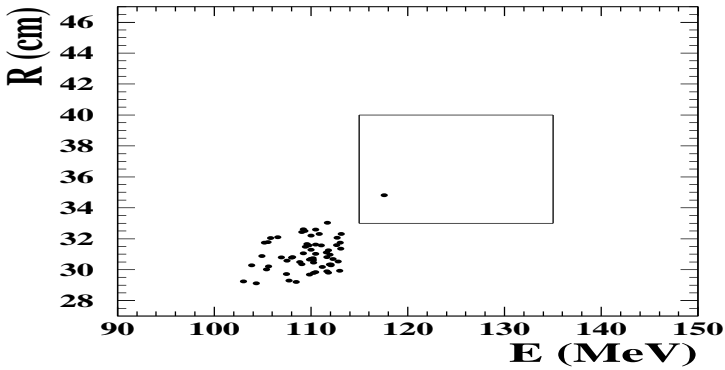,height=2.in,width=3.75in,angle=0}
\caption{Final E787 data sample from the 1995--97 data set after all cuts. One
clean \kpnn event is seen in the box. The remaining events are \kpp
background.
\label{fig:e787}}
\end{figure}

The E787 experiment at BNL has recently published an analysis of the 1995--97 
data sample~\cite{e787_pnn}: one clean \kpnn event
lies in the signal box (see Fig.~\ref{fig:e787}),
with a measured background of $0.08\pm0.02$ events.  Based on this one
event the branching ratio is B(\kpnn) = $1.5^{+3.4}_{-1.2} \times
10^{-10}$. From this measurement, a limit of $0.002 < |\vtd| < 0.04$
is determined; in addition, the following limits on $\lambda_t \equiv
V^*_{ts}\vtd$ can be set: $| Im(\lambda_t) | < 1.22\times10^{-3}$,
$-1.10\times10^{-3} < Re(\lambda_t) < 1.39\times10^{-3}$, and
$1.07\times10^{-4} < | \lambda_t | < 1.39\times10^{-3}$.  The final
sensitivity of the E787 experiment, based on data from 1995--98,
should reach a factor of two further, to the SM expectation for \kpnn.

A new experiment, E949, is under construction and will run in
2001--03. Taking advantage of the very large AGS proton flux and the
experience gained with the E787 detector, E949 with modest upgrades
should observe $\cal O$(10) SM events in a two year run. The background is
well-understood and is $\sim$10\% of the SM signal.

A proposal for a further factor of 10 improvement has been initiated
at FNAL. The CKM experiment (E905) plans to collect 100 SM events, with
$\sim$10 background events, in a two year run starting after 2005.  This
experiment will use a new technique, with K$^+$ decay-in-flight
and momentum/velocity spectrometers.

\boldmath
\subsubsection{\klpnn}
\unboldmath

The current best direct limit~\cite{e799_pnn} on \klpnn comes from the
KTeV run in 1997: B(\klpnn) $< 5.9\times10^{-7}$ (90\% CL).

An even more stringent limit can be derived in a model independent
way~\cite{grossman} from the E787 measurement of \kpnn:
\begin{eqnarray}
B(\klpnn) & < & 4.4 \times  B(\kpnn) \\ \nonumber
          & < &  2.6\times10^{-9} \; \; (90\%\,{\rm CL})
\label{eq:pnn}
\end{eqnarray}

The next generation of \klpnn experiments will start with E391a at
KEK, which hopes to achieve a sensitivity of $\sim10^{-10}$.  Although
the reach of E391a is not sufficient to observe a signal at the
standard model level, the experiment will be able to rule out large BSM
enhancements and learn more about how to do this difficult experiment.
This experiment would eventually move to the JHF and aim for a
sensitivity of ${\cal O} (10^{-14})$.

Two other experiments propose to reach sensitivities of ${\cal O}
(10^{-13})$: E926 (KOPIO/RSVP) at BNL and E804 (KAMI) at FNAL. KAMI
plans to reuse the excellent CsI calorimeter from KTeV and to operate
at high kaon momentum to achieve good photon energy resolution and
efficiency.  It will take advantage of the large flux available from
the Main Injector.  KOPIO follows a different strategy; the kaon
center of mass will be reconstructed using a bunched proton beam and a
very low momentum K$_L$ beam.  This gives two independent criteria to
identify background: photon veto and kinematics --- allowing
background levels to be directly measured from the data --- and gives
further confidence in any observed signal. The necessary flux will by
obtained from the very high AGS proton current. The low energy beam
also substantially reduces backgrounds from neutrons and hyperons.
After three years of running, 65 SM events are expected with a S/B
$\ge$ 2:1.

\section{Conclusions and Future Prospects}

The unprecedented sensitivities of rare kaon decay experiments in
setting limits on LFV have constrained many extensions of the SM. The
discovery of \kpnn has opened the doors to measurements of the
unitarity triangle completely within the kaon system.  Significant
progress in the determination of the fundamental CKM parameters will
come from the generation of experiments that is now starting.
Comparison with the B-system will then over-constrain the triangle and
test the SM explanation of {\it CP} violation.

\section*{Acknowledgments}

I would like to thank members of several experiments for access to
data and for useful discussions, in particular, I would like to thank
Bill Molzon, Bob Tschirhart, Tony Barker and Hong Ma.  This work was
supported under U.S. Department of Energy contract
\#DE-AC02-98CH10886.

\def\Journal#1#2#3#4{{#1}{\bf #2}, #3 (#4)}

\def\ARNPS{{\em Ann. Rev. Nucl. Part. Sci. }}
\def\NCA{{\em Nuovo Cimento }}
\def\NIM{{\em Nucl. Instrum. Methods }}
\def\NIMA{{\em Nucl. Instrum. Methods} \bf A}
\def\NPB{{\em Nucl. Phys.} \bf B}
\def\PLB{{\em Phys. Lett.}  \bf B}
\def\PRL{{\em Phys. Rev. Lett. }}
\def\PRD{{\em Phys. Rev.} \bf D}
\def\ZPC{{\em Z. Phys.} \bf C}
\def\PTP{{\em  Prog. Theor. Phys. }}


\begin{thebibliography}{99}
\bibitem{review2000} A.R.~Barker and S.H.~Kettell, \Journal{\ARNPS}{50}{249}{2000};
L.~Littenberg, {\it Proc. Rencontres de Moriond, Les Arcs, France, March, 2000};
S.~Kettell, {\it Proc.\ 3rd Int.\ Conf.\ B Phys.\ and CP Violation, Taipei, Taiwan, Dec.\ 1999},
also hep-ex/0002011;
W.~Molzon, {\it Proc.\ XIX Int.\ Symp.\ Lepton and Photon Interact., Stanford, August 1999.},
also  hep-ex/0001024;
{\it Proc.\ Chicago Conf.\ Kaon Phys., June  1999}, also http://hep.uchicago.edu/kaon99/.
\bibitem{kobayashi} N.~Cabibbo, \Journal{\PRL}{10}{531}{1963}; M.~Kobayashi, T.~Maskawa \Journal{\PTP}{46}{652}{1973}.
\bibitem{e871_me} D.~Ambrose, {\it et al.}, \Journal{\PRL}{81}{5734}{1998}.
\bibitem{e865_pme}R.~Appel, {\it et al.}, {\em Phys. Rev. Lett. } In press; also hep-ex/0005016 (2000).
\bibitem{e799_pme}K.~Arisaka, {\it et al.}, \Journal{\PLB}{432}{230}{1998}.
\bibitem{dambrosio}G.~D'Ambrosio, {\em et al.},  \Journal{\PLB}{423}{385}{1998}.
\bibitem{dumm} D.G.~Dumm and A.~Pich, \Journal{\PRL}{80}{4633}{1998}.
\bibitem{valencia} G.~Valencia, \Journal{\NPB}{517}{339}{1998}.
\bibitem{derafael} M.~Knecht, {\it et al.}, \Journal{\PRL}{83}{5230}{1999}.
\bibitem{donoghue}J.F.~Donoghue, F.~Gabbiani,  \Journal{\PRD}{51}{2187}{1995}.
\bibitem{glashow} S.L.~Glashow, J.~Iliopoulos, L.~Maiani, \Journal{\PRD}{2}{1285}{1970};
M.K.~Gaillard, B.W.~Lee \Journal{\PRD}{10}{897}{1974}.
\bibitem{e871_mm} D.~Ambrose, {\it et al.}, \Journal{\PRL}{84}{1389}{2000}.
\bibitem{wolfenstein} L.~Wolfenstein, \Journal{\PRL}{51}{1945}{1983}.
\bibitem{buras2} A.~Buras,  {\it et al.}, \Journal{\NPB}{566}{3}{2000}.
\bibitem{e871_ee} D.~Ambrose, {\it et al.}, \Journal{\PRL}{81}{4301}{1998}.
\bibitem{greenlee} H.B.~Greenlee, \Journal{\PRD}{42}{3724}{1990}.
\bibitem{e799_pee} T.~Yamanaka, {\it Proc. Rencontres de Moriond, Les Arcs, France, March 1999}.
\bibitem{e799_pmm} A.~Alavi-Harati, {\it et al.}, \Journal{\PRL}{84}{5279}{2000}.
\bibitem{daphne} S.~Kettell, {\it Proc.\ 3rd DA$\Phi$NE Work. Phys.\ and Det., Frascati, Italy, Nov.\ 1999}, also hep-ex/0002009.
\bibitem{sinb} G.~Bucahalla and A.~Buras, \Journal{\PLB}{333}{221}{1994};
G.~Bucahalla and A.~Buras, \Journal{\PRD}{54}{6782}{1996};
Y.~Nir and M.P.~Worah, \Journal{\PLB}{423}{319}{1998};
S.~Bergmann and G.~Perez, hep-ph/0007170.
\bibitem{grossman} Y.~Grossman and Y.~Nir, \Journal{\PLB}{398}{163}{1997}.
\bibitem{bb3} G.~Bucahalla and A.~Buras, \Journal{\NPB}{548}{309}{1999}.\label{ref:bb3}
\bibitem{jarlskog}C.~Jarlskog, \Journal{\PRL}{55}{1039}{1985}.
\bibitem{bb1} G.~Buchalla and A.~Buras, \Journal{\NPB}{412}{106}{1994}.
\bibitem{marciano} W.J.~Marciano and Z.~Parsa, \Journal{\PRD}{53}{R1}{1996}.
\bibitem{bb2} G.~Bucahalla and A.~Buras, \Journal{\PRD}{57}{216}{1998}.
\bibitem{buras} A.~Buras, hep-ph/9806471;  T.~Inami and C.S.~Lim, \Journal{\PTP}{65}{297}{1981}.
\bibitem{e787_pnn} S.~Adler, {\it et al.}, \Journal{\PRL}{79}{2204}{1997};
 \Journal{\PRL}{84}{3768}{2000}.
\bibitem{e799_pnn} A.~Alavi-Harati, {\it et al.}, \Journal{\PRD}{61}{072006}{2000}.
\end{thebibliography}
\end{document}